\documentclass[RNAAS]{aastex63}

\usepackage{xspace}

\newcommand{\name}{ATLAS18qtd\xspace}
\newcommand{\Ha}{H$\alpha$\xspace}
\newcommand{\LHa}{$L_{\rm H\alpha}$\xspace}

\shorttitle{ATLAS18qtd}
\shortauthors{Tucker \& Shappee}
\graphicspath{{./}{figures/}}

\begin{document}

\title{H$\alpha$ Luminosity of ATLAS18qtd Does Not Plateau in the Nebular Phase}

\correspondingauthor{Michael Tucker}
\email{tuckerma95@gmail.com}

\author[0000-0002-2471-8442]{Michael A. Tucker}
\affiliation{
Institute for Astronomy, \\
University of Hawai'i at Manoa \\
2680 Woodlawn Dr. \\
Honolulu, HI 96822, USA}
\affiliation{DOE CSGF Fellow}

\author{Benjamin J. Shappee}
\affiliation{
Institute for Astronomy, \\
University of Hawai'i at Manoa \\
2680 Woodlawn Dr. \\
Honolulu, HI 96822, USA}



\begin{abstract}

We present new spectroscopic and photometric observations of ATLAS18qtd/SN~2018cqj, a fast-declining Type Ia supernova with variable \Ha emission in previously-published nebular phase spectra. ATLAS18qtd is undetected in both spectroscopic and photometric observations which occurred at $\sim 540~\rm d$ after maximum light and $\sim 230~\rm d$ after the last \Ha detection. With these new non-detections, we place an upper limit on the \Ha luminosity of $\lesssim 1.1\times 10^{36}~\rm{erg}~\rm s^{-1}$ indicating the \Ha flux decreased by a factor of $\gtrsim 4$ since the previous detection. This upper limit excludes \Ha emission that plateaus or increases since the previous detection but cannot confirm that the \Ha emission decay rate is equivalent to the supernova decay rate.
\end{abstract}



\section{Introduction} \label{sec:intro}

Type Ia supernovae (SNe Ia) are key cosmological probes \citep[e.g., ][]{riess98, perlmutter99}, yet their progenitor systems are still widely debated. The two main theories for SNe Ia progenitors are the single- and double-degenerate scenarios, where `single' and `double' refer to the number of degenerate objects (i.e., white dwarfs, hereafter WDs) in the system (see \citealp{jha19} for a review). While searches for observational signatures from the single-degenerate scenario have mostly returned non-detections, the double-degenerate scenario struggles to explain SNe Ia interacting with nearby circumstellar material (CSM), referred to as SNe Ia-CSM, \citep{silverman13}. 

\name (SN~2018cqj) is a faint, fast-declining SN Ia \citep{prieto20} in IC~550 ($z=0.0165$) discovered by the Asteroid Terrestrial-impact Last Alert System \citep[ATLAS, ][]{tonry19}. Variable \Ha emission in nebular-phase spectra was discovered by \citet{prieto20} and the uncertain origin of the \Ha emission prompted our follow-up observations.

\section{Optical Imaging and Spectroscopy}\label{sec:methods}

\begin{figure*}
    \centering
    \includegraphics[width=0.95\linewidth]{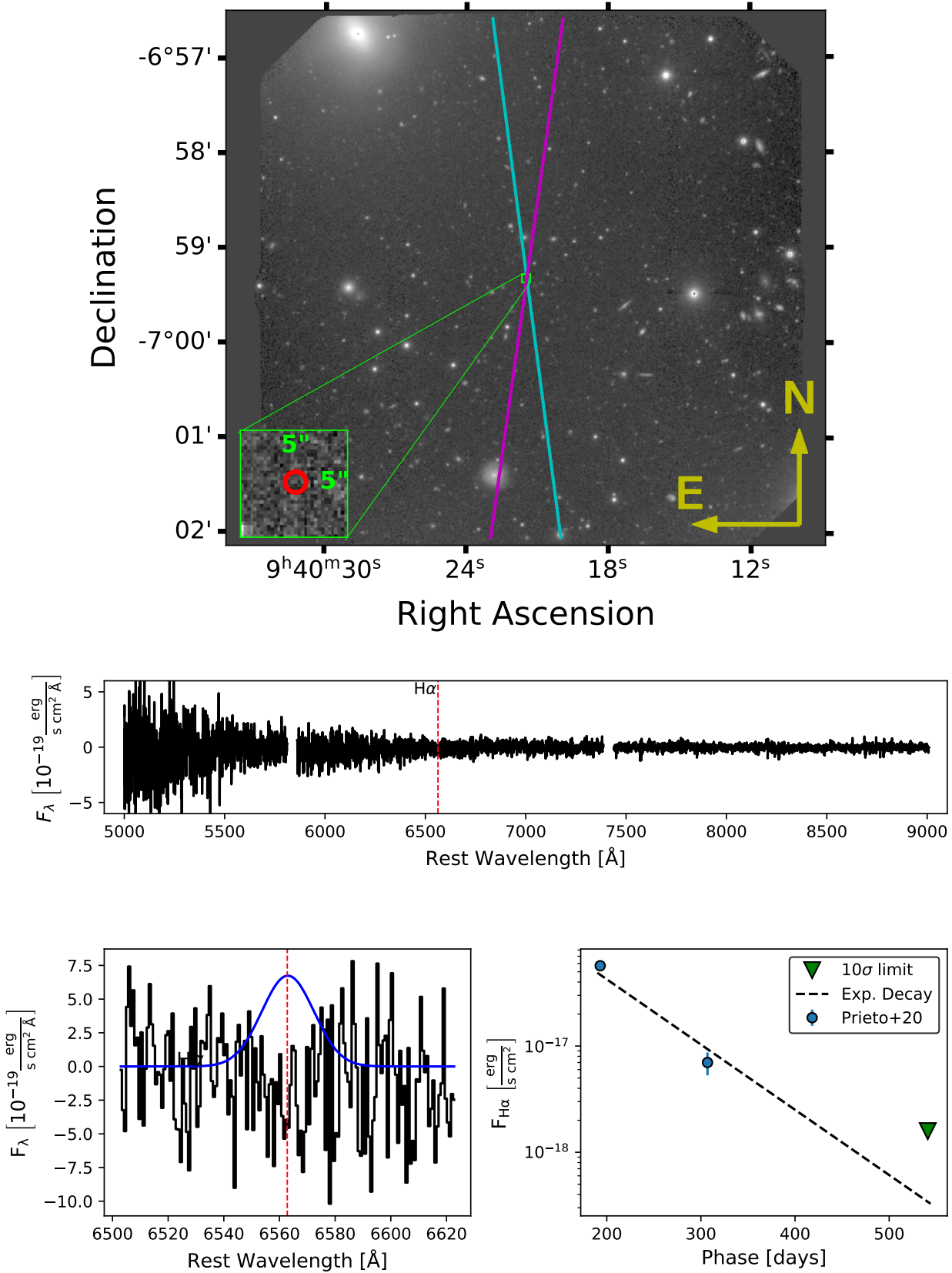}
    \caption{\textit{Top panel}: $r$-band image of \name. Thin rectangles represent 1\arcsec-wide slit orientations from each night when observations were conducted (cyan=Dec. 27; magenta=Dec. 28; 2019 UT). 
    \textit{Middle panel}: Non-detection spectrum of \name at $\sim 540$~days after maximum. 
    \textit{Bottom left panel}: Spectral region around \Ha showing no evident emission and our $10\sigma$ flux limit in blue.
    \textit{Bottom right panel}: \Ha evolution including our new $10\sigma$ upper limit. The dashed black line represents the expected flux if the \Ha emission decays at the same rate as the $V$-band light curve. 
    }
    \label{fig:1}
\end{figure*}

New optical spectroscopy and $r$-band imaging data were acquired on Dec 27 \& 28, 2019 UT with the Gemini Multi-Object Spectrograph (GMOS; \citealp{GMOSref}). Initial data reduction procedures roughly follow the Gemini Data Reduction Cookbook\footnote{\url{http://ast.noao.edu/sites/default/files/GMOS_Cookbook/}}. Additionally, we implement \textsc{lacosmic} \citep{lacosmicref} to reject cosmic rays. 

The $r$-band image has a total exposure time of $\approx 1200~\rm s$ and has a measured FWHM of $\approx 0.75^{\prime\prime} (=4.6~\rm{pixels})$.  The final WCS solution is verified with 162 PanSTARRS objects and has an RMS of $\sim 1.5$~pixels. We compute aperture photometry of stars in the image and derive a photometric zeropoint of $32.71\pm0.03$~mag. With this photometric zeropoint, we place a $3\sigma$ upper limit at the location of \name of $r > 25.2$~mag.

The spectroscopic observations total 8100~s of exposure time split across 2019 Dec. 28 \& 29 UT. On both nights nearby objects fall within the slit and are used to anchor the trace location for \name. We extract the non-detection spectrum for each observation and provide the final spectrum in Fig. \ref{fig:1}. A spectrophotometric standard star was used to correct the extracted spectrum for instrumental throughput and atmospheric attenuation. 

\name is undetected in both the photometric and spectroscopic observations, yet the spectrum must be on an absolute flux scale to place a useful limit on \Ha emission. Three slit-coincident point sources are used to estimate the reliability of the flux scale from the spectrophotometric standard star. After correcting for slit losses we calculate synthetic magnitudes from the flux-calibrated spectra and compare these results to magnitudes derived from the $r$-band image. The flux calibration is accurate to within a factor of $\sim 2$ without additional corrections, attributable to the excellent observing conditions (mean airmass $<$ 1.2, seeing $\sim$ 0.8$^{\prime\prime}$, and little-to-no cloud cover). 

\section{Results}\label{sec:discuss}

The new observations occurred $\approx 541$~days after maximum light using the $t_{\rm{max}}$ from \citet{prieto20}. Utilizing Eq. 3 from \citet{tucker2020}, the $10\sigma$ upper limit on \Ha emission is $F_{\rm H\alpha}(10\sigma) < 1.6\times10^{-18}~\rm{erg}~\rm s^{-1}~\rm{cm}^{-2}$, or equivalently $L_{\rm H\alpha} (+550\rm d) < 1.1\times10^{36}~\rm{erg}~\rm s^{-1}$ (Fig. \ref{fig:1}). The conservative $10\sigma$ upper limit is chosen to better reflect the inherent uncertainties in our analysis, such as imprecise flux calibration and neglecting any uncertainty in the redshift-derived distance.

The nebular spectra of \name at +193 and +307~days after maximum have measured \Ha luminosities of $L_{\rm H\alpha}(+193\rm d) = 3.8\times10^{37}\rm{erg}~\rm s^{-1}$ and $L_{\rm H\alpha}(+307\rm d) = 4.6\times 10^{36}~\rm{erg}~\rm{s}^{-1}$ \citep{prieto20}. Additionally, the nebular $V$-band decline rate for \name was measured to be $\sim 0.015~\rm{mag}~\rm{day}^{-1}$. If $L_{\rm H\alpha}$ decays similarly to the SN bolometric luminosity, which is in turn nearly proportional to the $V$-band decline rate \citep[e.g., ][]{shappee17,dimitriadis17}, the predicted \Ha luminosity during our spectroscopic observations is $L_{\rm H\alpha}(+550~\rm d)\sim 3\times10^{35}~\rm{erg}~\rm{s}^{-1}$, a factor of $\approx3.5$ lower than our \Ha upper limit (Fig. \ref{fig:1}).

Our new upper limit on \Ha constrains some CSM scenarios. SNe Ia-CSM typically exhibit \Ha luminosities that plateau in the nebular phase and the emission lasts out to 500~days after maximum (e.g., PTF11kx, \citealp{silverman11kx}) although observations at these epochs are scarce. Any continuation or increase in \LHa would be detectable in our spectrum, indicating the \Ha must decline by a factor of $\gtrsim 4$ between the prior spectrum at +307~d and our later observations. Thus, we can place a lower limit on the \Ha decline rate of $\gtrsim 0.007$~mag/day. This limit is consistent with the $V$-band decline rate of $\sim 0.015$~mag/day but cannot confirm that these decay rates are equivalent. Another possible explanation for the \Ha emission that does not invoke the presence of CSM is material stripped from a companion star during the explosion \citep[e.g., ][]{marietta00}. However, the inferred stripped mass values are an order of magnitude lower than expected from models in the literature \citep{prieto20, dessart20}. 

\section*{Acknowledgments}
We thank Aaron Do for assisting with the SNIFS/UH88 acquisition imaging. M.A.T. acknowledges support from the DOE CSGF through grant DE-SC0019323. B.J.S. is supported by NSF grants AST-1908952, AST-1920392, and AST-1911074.

\bibliography{sample63}{}
\bibliographystyle{aasjournal}

\end{document}